\documentclass[12pt]{IEEEtran}
\IEEEoverridecommandlockouts
\usepackage{cite}
\usepackage{amsmath,amssymb,amsfonts}
\usepackage{algorithm}
\usepackage{algorithmic}
\usepackage{graphicx}
\usepackage{caption}
\usepackage{textcomp}
\usepackage{xcolor}
\usepackage{url}
\usepackage{booktabs}
\usepackage{siunitx}
\usepackage{comment}

\def\BibTeX{{\rm B\kern-.05em{\sc i\kern-.025em b}\kern-.08em
    T\kern-.1667em\lower.7ex\hbox{E}\kern-.125emX}}
\begin{document}

\title{Distributed Learning Meets 6G: A Communication and Computing Perspective
}

\author{\IEEEauthorblockN{Shashank Jere, Yifei Song, Yang Yi and Lingjia Liu}

\thanks{The authors are with Wireless@Virginia Tech, Bradley Department of Electrical and Computer Engineering, Virginia Tech, Blacksburg, VA, USA. The work is supported by the US National Science Foundation (NSF) under grants ECCS-1811497, CNS-1811720, and CCF-1937487. The corresponding author is L. Liu (ljliu@ieee.org). This article has been accepted to \emph{IEEE} Wireless Communications Magazine under the Series ``AI-Powered Telco Network Automation: 5G Evolution and 6G''.}

}

\maketitle

\begin{abstract}
With the ever-improving computing capabilities and storage capacities of mobile devices in line with evolving telecommunication network paradigms, there has been an explosion of research interest towards exploring Distributed Learning (DL) frameworks to realize stringent key performance indicators (KPIs) that are expected in next-generation/6G cellular networks. 
In conjunction with Edge Computing, Federated Learning (FL) has emerged as the DL architecture of choice in prominent wireless applications. 
This article lays an outline of how DL in general and FL-based strategies specifically can contribute towards realizing a part of the 6G vision and strike a balance between communication and computing constraints.
As a practical use case, we apply Multi-Agent Reinforcement Learning (MARL) within the FL framework to the Dynamic Spectrum Access (DSA) problem and present preliminary evaluation results.
Top contemporary challenges in applying DL approaches to 6G networks are also highlighted.  
\end{abstract}


\section{Introduction}
\label{introduction}
The past three decades have witnessed an evolution of the telecommunications industry from 2G to 5G, each enabling richer user experiences compared to its previous generation, backed by increasingly sophisticated advancements in the air interface and the core network. Even though commercial 5G non-standalone (NSA) networks were first launched in 2019, it is fair to say that they are yet to realize their full potential across a wide range of applications ranging from enhanced mobile broadband (eMBB) to massive machine-type communications (mMTC) and internet of things (IoT) to ultra-reliable low-latency communications (uRLLC), as originally envisioned for 5G. 
Looking at the journey of successive generations of commercial mobile networks, every generation has required approximately a decade of focused research for successful large-scale deployment based on that technology. With 5G standalone (SA) networks too being planned for roll-out in 2023, it is not unreasonable to expect the launch of the first 6G network by 2030. Therefore, now is the right time to ask the question: \emph{What will 6G look like?}


\section{A Vision for 6G}
\label{6g_vision}
The answer to what 6G should look like is multi-dimensional and a truly forward-looking vision for 6G must consider not only futuristic enabling technologies and services, but also exploit the full spectrum of emerging value chains and upcoming business verticals. The first deployed 6G network can be expected to be an evolution of 5G networks to some degree but also to simultaneously incorporate radically disruptive technologies for realizing use cases that are currently not part of 5G and which may not be fully realizable with existing technologies.

For 6G to be the ubiquitous wireless network of choice, it is imperative that it empower mobile network operators to deliver wide-ranging services to not just the final consumers (UEs), but also to the larger enterprise services, which can be broadly categorized into the Business-to-Business (B2B) services category, which includes logistics, manufacturing, transportation, health, banking and government sectors amongst others. While 5G has made an effort to serve a subset of such B2B verticals via mMTC and uRLLC, a significant effort has to be made to define clear KPIs in 6G for different verticals within the B2B category. An always-connected end user which may be a consumer in a Business-to-Consumer (B2C) service or a business (B2B), must be able to access all available digital services by harnessing the various well-defined capabilities of the ubiquitous 6G network. To this end, we present a sampling of 6G use cases in this section. This list is not exhaustive but includes a collection of verticals that could be implemented in 6G based on existing KPI specifications currently available in part in 5G standards. 
\subsection{Manufacturing}
Heavy industry, especially the manufacturing sector relies on high-precision equipment that often has to function in cooperation with each other, e.g., robotics in the automobile or the parts manufacturing industry. 
In the 6G vision of the connected industrial floor, it is expected that multiple radio access technologies (RATs), together with time-sensitive networking will deliver the reliability and latency performance that would support various industrial applications.
A preliminary version of this is espoused as part of the Industry 4.0 vision, however 5G New Radio (NR) standards aimed at Industry 4.0 are not sufficient to realize the expanded use cases of the next-generation industrial floor. As an example, automated and real-time monitoring of critical infrastructure such as power grids and energy supply lines demands uRLLC links involving dual mobility of robots and human workers. 

\subsection{Healthcare}
The COVID-19 pandemic has demonstrated the importance of healthcare services being delivered directly to the patient at their homes, and the need of such services to be supported in 6G. Transitioning to home care also reduces operational and administrative costs for hospitals and caters better to high-risk patients with potential mobility impairments. Remote patient monitoring can be enhanced using Augmented Reality (AR) and Virtual Reality (VR). High data rate and extremely high-reliability links with latency requirements of less than a millisecond will be the need for robot-assisted telesurgeries. Another key enabling technology to make telemedicine possible would be real-time tactile feedback, which has not been achieved yet. However, 6G which will likely ensure higher spectrum usage together with edge-assisted distributed learning techniques can potentially deliver the stringent KPIs needed in telemedicine.

\subsection{Public Services and Safety}
\label{public_safety}
Public Safety (PS) operations are critical in dispensing critical and in some cases, life-saving information to citizens from government agencies. First responders are the most important component in the complete chain of command and traditionally have relied on unreliable voice-only links. However, upgraded capabilities such as high-definition real-time streaming video from body cameras, and real-time access to sensor data including thermal sensors will enhance the capabilities of first responders for improved crisis mitigation. 
Device-to-Device (D2D) communications including remote robot control for applications such as bomb defusal and operating robots in incident locations unsuitable for humans will also require KPIs that are currently not met by 5G.
In order to meet these in 6G, there must be a focus on improved coverage and the ability to support a large number of connections in a dense environment. The PS networking infrastructure in 6G must also include KPIs that ensure efficient usage of battery-operated end-user devices when receiving PS messages.

\section{Communication and Computing Tradeoff in 6G}
Early works that played a key part in the ideation of 5G emphasized the need to transition towards a software-centric approach starting from the network core to the air interface. Software-defined Networking (SDN), which marks a shift from the traditional hardware-centric approach along with Network Function Virtualization (NFV), will continue to be the primary enabler in 6G networks too, as they have been in 5G.
In parallel, the overall Mobile Edge Computing (MEC) paradigm advocates for the re-structuring of radio access network (RAN) and core network functions by transferring some of the Base Station (BS) functionality upstream to the cloud and transferring some of the core network functionality downstream. This creates a clearly identifiable ``edge'' or ``fog'' domain between the BS and the end device.

Although cloud computing~\cite{haber2022cloud} has brought richer and more complex applications to end users by harnessing the power of the remote cloud server, extremely sensitive latency requirements specified for use cases in 5G and potentially in 6G have demanded an alternate approach.
Due to the complex traffic distributions in modern wireless networks, the network architecture is becoming increasingly heterogeneous.
There are multiple types of network access nodes providing reliable and seamless connectivity for mobile users such as a macro BS, a small cell BS, and a WiFi access point (AP), to name a few.
These network access nodes support edge computing at network edges with low transmission latency.
Due to the different characteristics of network access nodes such as coverage ability and transmit power, the design of the co-existence of heterogeneous MEC networks has attracted increasing attention~\cite{cao2020overview}.
The cooperative computational offloading between multiple network access nodes needs to be well-designed.

Under such a heterogeneous network architecture, intelligent task allocation and resource allocation among different network nodes can significantly improve system performance. 
On one hand, cooperation between the edge and the cloud can be achieved to enhance the quality of service (QoS) of IoT tasks further. Specifically, cloud servers can process tasks that require intensive computation, while edge servers can process tasks that work with a small data size or have a low latency requirement. On the other hand, intelligent task allocation among edge servers can effectively offload tasks from overloaded edge servers to underutilized ones, and thus reduce the execution delay of tasks~\cite{cao2020overview}.

\section{Distributed Learning for 6G}
Due to the dispersed and occasionally sparse nature of cellular wireless networks consisting of possibly heterogeneous end devices, the distributed learning (DL) paradigm has emerged as being vital in applying ML approaches to wireless network problems in general. The factors that make DL fit for application to wireless networks are multi-fold:
\begin{itemize}
    \item As mobile and IoT devices become computationally more capable with higher storage capacities, they will also generate exponentially large amounts of local user data and contextual sensing data originating from diverse applications.
    
    \item Due to constraints of sending large amounts of data from end devices over bandwidth-limited wireless channels to server nodes and due to user data privacy concerns, it is not optimum to send local data to the server node (aggregator) in every training round. 
\end{itemize}
Therefore, it is beneficial for the end devices to generate and store locally generated data on-device and only transfer the model parameter updates obtained from local training to the central server, which could be used to update a global ML model. This is referred to as the ``parameter server'' architecture which can be categorized as a centralized multi-node distributed ML approach. Federated Learning (FL) is one of the most popular parameter server architecture variants and makes up the vast majority of distributed ML research in wireless communication systems.
There exist other decentralized DL approaches such as MapReduce~\cite{rajendran2021mapreduce}, AllReduce, and All-to-All among others. However, they are not widely applied to wireless networks due to practical bandwidth and latency constraints, and hence our focus in this article will be on the FL architecture and its associated algorithms that can be applied to wireless networked systems.

FL lends itself particularly well to application in large-scale wireless networks such as cellular systems~\cite{JereEdgeComm2022}. In particular, FL addresses the privacy concerns of heterogeneous users that are not well-addressed in more conventional DL architectures which may involve sharing of local user data with the central server or with each other. Additionally, since FL only requires sharing of parameter updates from the participating devices to the aggregator and not the local data itself, FL reduces the overall communication overhead~\cite{JereEdgeComm2022} and can tackle wireless channel uncertainties more effectively, thereby improving reliability.


\subsection{Federated Learning Preliminaries}
A traditional cloud-only-based ML approach offloads data sensed at end devices to the remote cloud server for centralized training in order to train a common model for future inference. However, the training time in the cloud may be impractical owing to the large volume of the sensed data that needs to be utilized in the training process, and in part due to the training computational complexity if the ML model being significantly large. Meanwhile, as the cloud server may be physically distant from the end devices, these devices may suffer from large communication delays. To solve this problem, FL facilitated by MEC can be a promising approach to shift from a centralized training paradigm to a more practical decentralized training one. Federated Learning~\cite{bonawitz2017practical} enables aggregation of the ML models on different end devices which are trained using their local datasets and cooperatively learning the global model. 
Specifically, at the beginning of each round, the server sends the current global model to each participating end device. The end devices (clients) then train their individual local models based on their own limited datasets and transfer back the model parameter updates at the end of each training round to an aggregator at the server. 
This can be repeated for as many training rounds as necessary for the global model to achieve the desired accuracy. 
FL distinguishes itself from other distributed learning schemes owing to certain unique factors. Firstly, the assumption that the data samples sensed at the different end devices are realizations of independent and identically distributed (i.i.d.) random variables may not hold in FL, since the local dataset of a single user's end device may not be representative of the overall population distribution. Secondly, the local datasets generated across federated learners may differ greatly in size, causing an imbalanced distribution. 
This imbalance in dataset sizes is primarily due to the different types of IoT devices (e.g., smartphone or vehicle) and different application scenarios (e.g., a maps application on a smartphone may generate more data for an active city user than a less active rural user). 
Thirdly, in the FL setting, the total amount of sensed data samples contributing towards learning the global model at the edge server is much larger than that available for local training at each user. 
Finally, most federated learners are mobile devices (e.g., smartphones, wearables, drones, vehicles, etc.) with possibly unreliable wireless connectivity to the FL edge server. This implies that the aggregator may have to support offline learners or learners with slow connectivity, especially in the cellular uplink scenario. In the context of these differentiating factors, FL provides clear advantages in wireless applications that may not be available in other decentralized ML approaches.


\subsection{Case Study: FL for Dynamic Spectrum Access}
In this section, we consider Dynamic Spectrum Access (DSA) as a special application in which FL can be applied for superior performance. First, we introduce some preliminaries on DSA.
To efficiently utilize spectrum resources, two types of spectrum management mechanisms can be utilized: static and dynamic.
Static spectrum sharing groups and reorders all spectrum resources to assign the same portion back to Service Providers (SPs). 
The licensed SPs schedule these spectrum resources to their subscribers accordingly.
On the other hand in dynamic spectrum access, the spectrum resources are dynamically allocated to both licensed SPs and unlicensed SPs with and without a QoS guarantee respectively. This provides an efficient way of utilizing available radio resources and alleviating spectrum shortages without adding new spectrum resources for unlicensed SPs. 
Licensed users and Unlicensed users are referred to as Primary Users (PUs) and secondary users (SUs) respectively henceforth in this article.


ML methods have been used previously in DSA applications to allocate spectrum resources more effectively. 
For example, Deep Reinforcement Learning (DRL) was introduced in~\cite{RubayetAI} as a natural tool for dynamic spectrum access and sharing.
Specifically, the DRL agent takes an action based on an observation of the environment, receives a reward from the environment depending upon the action taken, and then transitions into a new state. The goal of the DRL agent is to find a policy that optimizes the cumulative reward.
The DRL framework considered in this work is Multi-Agent Reinforcement Learning (MARL). 
Here, multiple agents are involved in the system thereby transforming this into an optimization problem that incorporates the policies of all agents involved. 
The individual actions, rewards, and states of every agent impact that of every other agent~\cite{zhang2021multi}.
MARL allows these agents to communicate with the server and process their distributed tasks in parallel once the agents receive them. For example, agents can share their experiences with each other to boost their learning convergence. Furthermore, a MARL system allows the addition of new agents into the system and replacing inactive agents.
However, MARL suffers from a prohibitive computational time due to its exponential complexity which is a function of the problem's dimensionality. It is also affected by the environment's non-stationarity as well as the exploration and exploitation trade-off~\cite{zhang2021multi}. 
To mitigate these issues, solutions such as the Deep Q-Network (DQN)~\cite{zhang2022dqn}, its Reservoir Computing (RC) version known as the Deep Echo State Q-Network (DEQN) proposed in~\cite{DEQN}, etc. have been proposed. 
In what follows, we describe how MARL-enabled FL can be configured to tackle the DSA problem.

\subsubsection{Outline}
\begin{figure}[!t]
\includegraphics[width=\linewidth,clip]{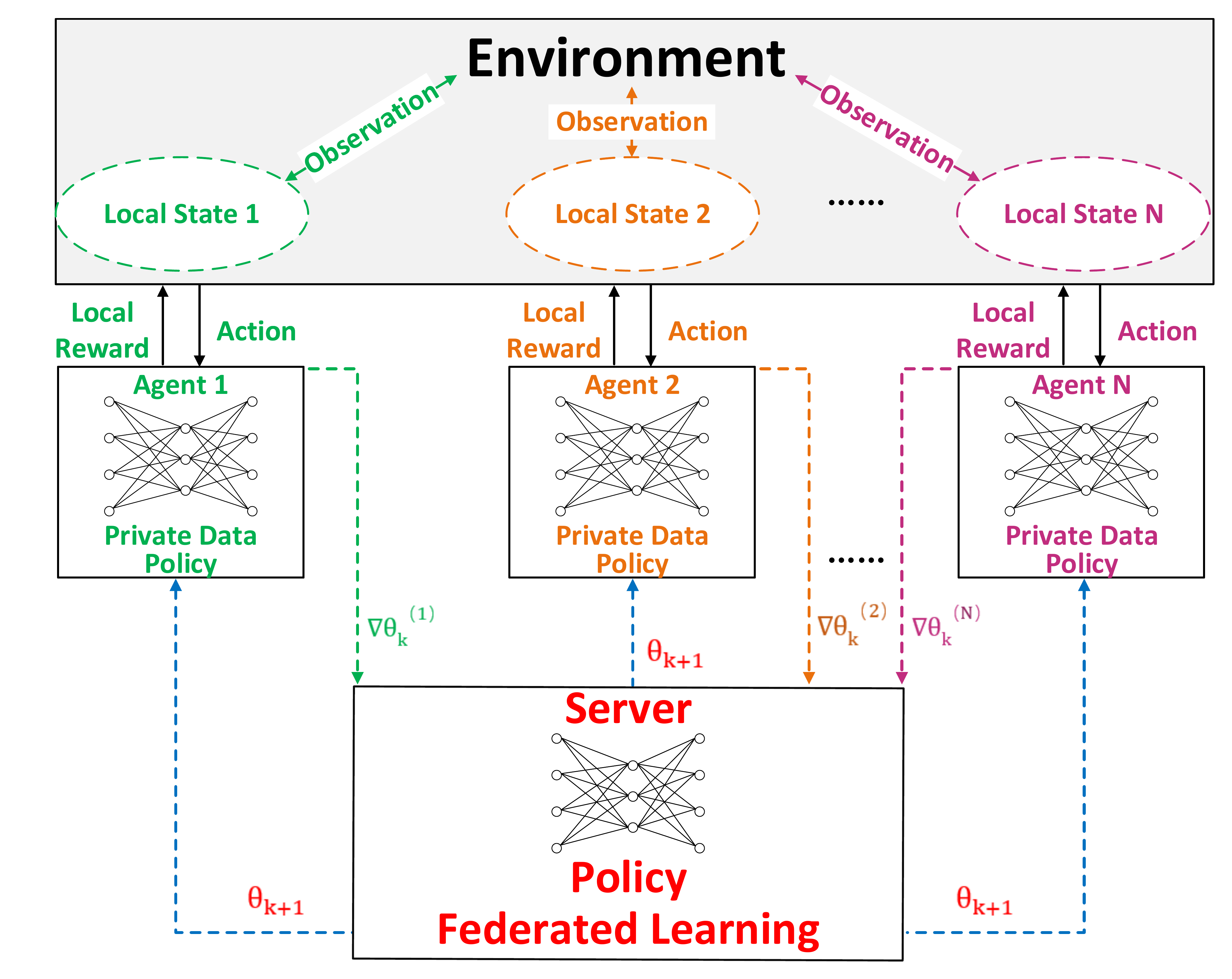}
\caption{Multi-Agent Reinforcement Learning (MARL) in a Federated Learning framework.}
\label{FL_MARL}
\end{figure}

Existing MARL algorithms assume that a joint reward is received by all agents, or that each agent receives an individual reward but shares it with other agents.
However, this assumption may not be practical in certain real-world applications since agents may not share their observations and received rewards due to data privacy and security concerns.
In the MARL-enabled FL system being considered, the agents do not share their local observations and rewards with other agents but update their policies to maximize their individual long-term local rewards. 
The objective of this system is to optimize the joint long-term reward, expressed as the sum of each agent's long-term local reward. The architecture of FL in a MARL setting can be depicted in Fig.~\ref{FL_MARL}, where $\mathbf{\theta}_k$ represents the model parameters in communication round $k$ and $\nabla \theta_k^{(n)}$ represents the model parameters updates in round $k$ sent by user $n$ to the server. 
We select the Signal-to-Interference-plus-Noise Ratio (SINR) as our quality metric~\cite{DEQN}, which takes into account all the factors affecting the SUs in the network, such as the receiver thermal noise, the BS transmission power and the interference between potentially simultaneously transmitting BS-user pairs. 
The user throughput is used as each user or agent's local reward function. 
Since MARL enables user interaction with the environment and training a shared model for maximizing a long-term reward, it aligns well with the FL idea of using a shared global model amongst all users.
This also presents an example of how FL can be applied to accommodate a large number of SUs in the Public Services and Safety use case for 6G which was outlined in Sec.~\ref{public_safety}.

\subsubsection{Spectrum Access Policy}
We model an SU's spectrum access strategy to utilize the spectrum resources efficiently as follows. There are $N$ SUs and $M$ channels with $(N>M)$, so that each SU can only access one channel at a specific time. To avoid interference from unlicensed users, SUs are not allowed to transmit on a particular channel when a PU is occupying that channel. 
However, an SU may interfere with another SU. The channel access activity of the PUs is modeled as a Markov Process.
To collaboratively avoid interference among SUs, we apply the previously outlined spectrum access framework to SUs.

We use a decentralized policy gradient method in our MARL system to optimize the joint reward. An initialized policy network is first distributed to all agents. The policy network which is implemented as a neural network at each agent, is updated based on its own observation of the environment. In each communication cycle, the agent empties its buffer, observes the environment, takes an action based on its policy, and receives a reward from the environment. After repeating the aforementioned steps for a sufficient number of iterations, each agent learns an updated local model. 
The updated local models are shared with and aggregated at a central server to update the global model.

%

%
\subsubsection{User Participation}
\label{Q_learning}
Selecting the appropriate number of users in each communication cycle of the FL training process is critical for accelerating convergence. To this end, we consider performing partial user participation during each round of aggregation in the FL algorithm. For a given number of participating users, we assume that the probability of a particular user being selected for participation in a specific training round is uniform. Only the local model weights of the participating devices are aggregated at the central server and only the participating devices in each aggregation cycle receive a model update under the proposed framework. Therefore, the RL agent deployed at each SU does not need to know about or depend on data samples from other SUs. 

\subsubsection{Simulation Results}
In our simulation setup, we randomly place $8$ BS-UE pairs in a $400$-m by $400$-m area and configure four different frequency bands as the available communication channels.
We only consider downlink, i.e., communication from the $8$ BSs to their respective UEs.
These $8$ BSs act as the SU transmitters (SU Tx) and the $8$ UEs act as the SU receivers (SU Rx). The probability of each of the $4$ channels being occupied by a PU is set to $20\%$.
At each time step $t$, the policy maps an agent's observation to its action such that the agent chooses to access one of the four channels or remain idle, after which it receives a reward.
The observation consists of the averaged historical throughput up to the previous time step and the throughput in the previous time step on all channels.
We compare our MARL-enabled FL with a traditional distributed learning approach, both deploying an RL agent that uses a two-layer feedforward neural network as its kernel. 
In the traditional DL setting, each SU Tx receives a model from the central server and starts updating its local model such that its local reward is maximized, without any further communication with the central server or with other SU Tx's. On the other hand, MARL-enabled FL can provide better overall performance at the system level since it enables indirect cooperation amongst SU Tx's via periodic aggregation of their individual local models at the central server.
With traditional DL, the users compete with each other for limited resources without cooperation. The users can take one of five actions, namely $[0, 1, 2, 3, 4]$, where $0$ indicates no channel access while any index from $1$ to $4$ represents accessing one of the four available channels. 
The local model at each agent is trained after $50$ time steps which is defined as one episode.
The global model in FL is aggregated after every $4^{\text{th}}$ episode. 
This periodic model aggregation and update helps with the long-term sum of rewards which is correlated with the user throughput. In other words, model aggregation allows users to ``peek'' into the environment of other users and enables user collaboration.

\begin{figure}[!t]
    \centering
    \includegraphics[width=1.0\linewidth, clip]{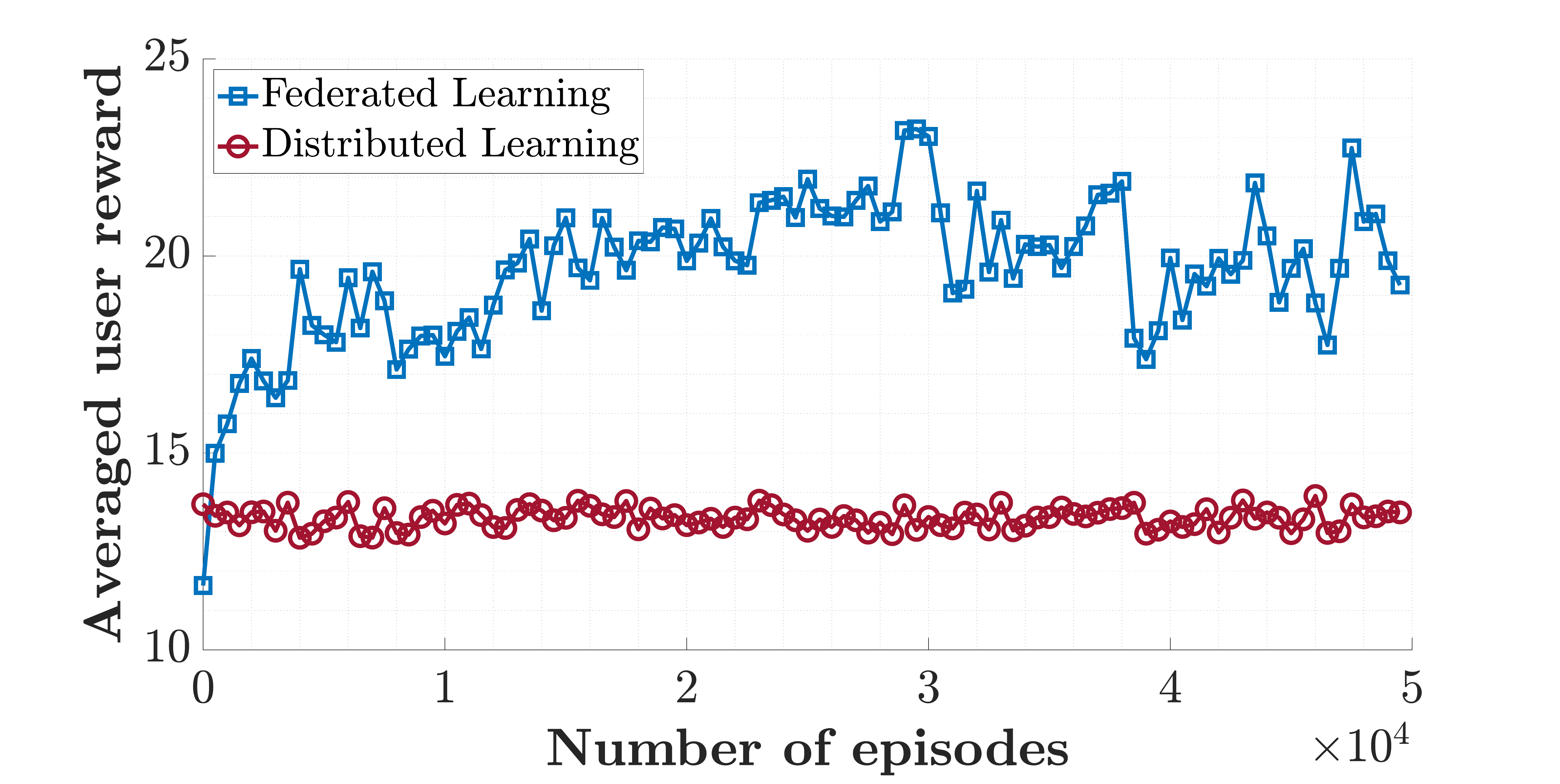}
    \caption{Averaged user reward in Federated Learning (FL) vs. conventional Distributed Learning (DL).}
    \label{FL_time}
\end{figure}

\begin{figure}[!t]
    \centering
    \includegraphics[width=1\linewidth, clip]{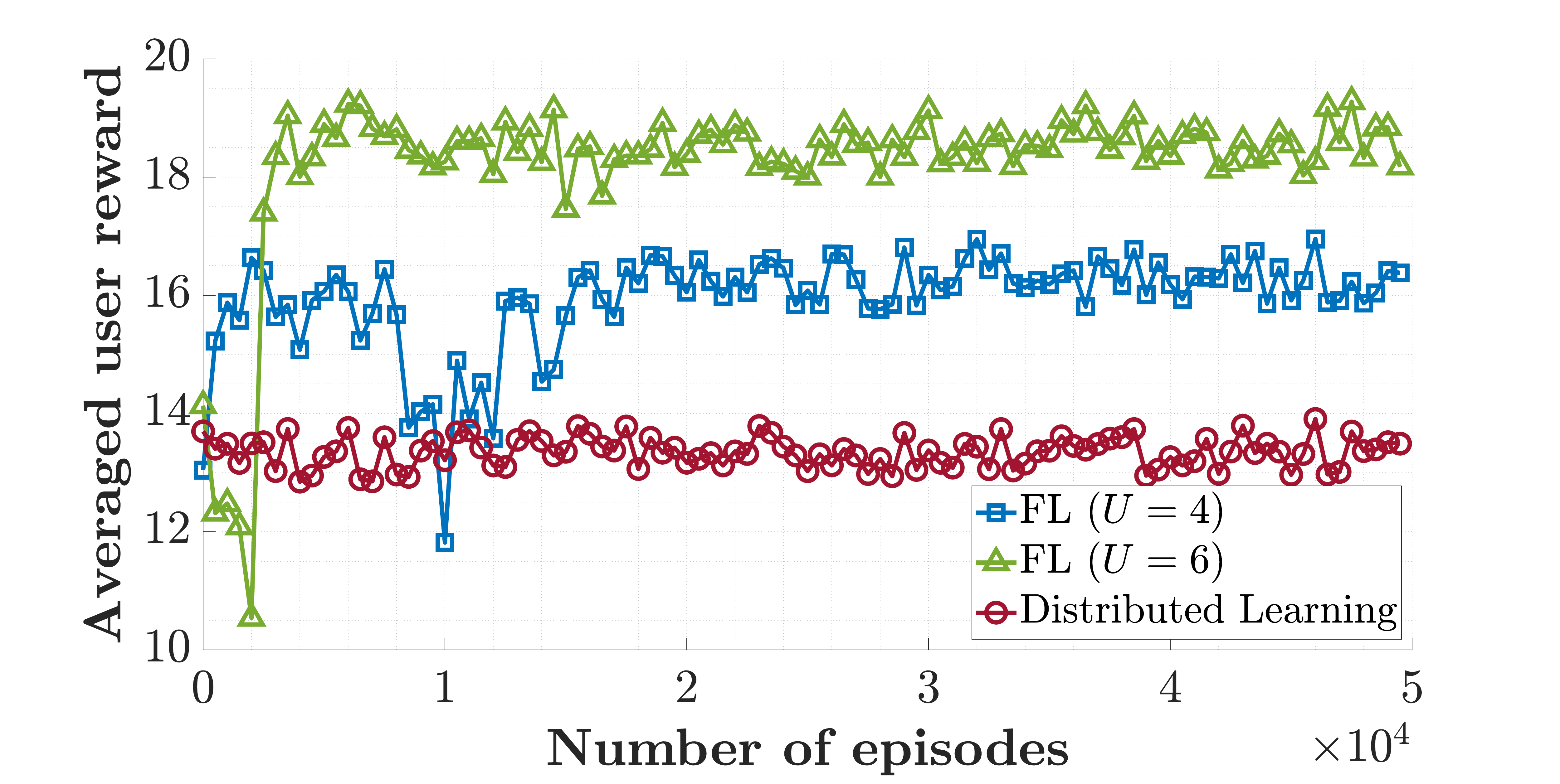}
    \caption{Averaged user reward vs. Number of participating users.}
    \label{Number_Users}
\end{figure}

Figure~\ref{FL_time} shows that MARL-enabled FL results in a higher (local) reward averaged across users and thus a higher joint reward as compared to DL, also implying a higher overall user throughput although FL requires a greater number of communication rounds between the SU Txs and the central server compared to conventional DL.
This points towards a trade-off between the communication overhead and the achievable system throughput using DL/FL methods.
Figure~\ref{Number_Users} investigates the FL framework for varying numbers of participating users $U$ $(<N)$ in every aggregation round, i.e., every $4^{\text{th}}$ episode. It shows that the greater the number of participating users, the higher the averaged user reward and thereby the system throughput. 
Meanwhile, this partial participation mechanism allows the framework to be more flexible in choosing users with better channel conditions, more relaxed energy constraints, and sufficient computational resources.
The complete set of parameters used in our simulation is summarized in Table~\ref{sim_params}.
The total number of episodes in the MARL algorithm is set to $50,000$.

\begin{table}[h]
	\centering
	\caption{Simulation Parameters.}
	\label{tab:parameters}  
	\begin{tabular}{p{0.22\textwidth}p{0.18\textwidth}}
	\hline
	\textbf{Parameter} & \textbf{Value}  \\ \hline
	Channel bandwidth  & $10$ MHz \\ \hline
	Path loss model & $41 + 22.7 \log_{10}(d[m])$ dB \\ \hline
	Small-scale fading & Rician distribution \\ \hline
	Line-of-Sight path coefficient & $5$ \\ \hline
	Noise spectral density  & $-174$ (dBm/Hz) \\ \hline
	Total episodes explored & $50000$  \\ \hline
	Time steps per episode  & $50$ \\ \hline
	Local learning rate  & $0.01$ \\ \hline
    Decay factor  & $0.9$ \\ \hline
    \end{tabular}
    \label{sim_params}
\end{table}

\section{Top Challenges for Distributed Learning in 6G}
The intricate balance between the remote cloud server and the edge node while providing end users with a high QoS that requires heavy computation but also adherence to extremely low latencies will remain a key challenge in most 6G use cases, one which DL and especially FL can potentially address.  
In this section, we provide a brief sampling of other related open problems that implementation of DL and specifically FL strategies for 6G will likely encounter. 

\subsection{Generalization}
One of the most prominent features in a distributed wireless network with a potentially large number of heterogeneous devices is the possibility of mobility of these devices, to the extent that sufficiently high-speed mobility for even a subset of devices may render the training and testing data distributions to be significantly different. While approaches such as domain adaptation can be used to improve the inference performance in the presence of such a training-test mismatch, implementing it on a large scale with acceptable on-device computational complexity is still an open problem. This applies not just to FL but to any DL approach in which statistically distinct training and test datasets are a possibility.

\subsection{Privacy Issues in FL} 
Maintaining the maximum possible number of participating devices in the FL training process is always a problem, especially in an unreliable wireless environment. Furthermore, to save energy, battery-powered IoT devices incorporate strategies suited to DL, e.g., opting out of certain training rounds. Although the assurance of local user data privacy is a standout feature of FL, malicious actors may still be able to glean critical system information from model changes~\cite{li2022soteriafl}. 
Although newer methods such as secure multiparty computation (SMC)~\cite{li2020privacy}, differential privacy~\cite{mohammadi2021differential} or secure aggregation~\cite{bonawitz2017practical} seek to improve the privacy of FL, these approaches generally sacrifice inference performance for privacy. Understanding and balancing these costs is a significant difficulty in implementing private FL systems, both theoretically and practically~\cite{FL_challenges_review}.

\subsection{Asynchronous FL Optimization} 
Although the synchronous FL model provides better convergence guarantees, it is sensitive to the Straggler effect~\cite{chen2020asynchronous}. The asynchronous FL model is more suitable in practice, especially when end devices differ in terms of hardware, network connection reliability, and battery capacity, resulting in substantial heterogeneity in system parameters throughout the network~\cite{FL_challenges_review}.
There needs to be a theoretical investigation into the convergence bounds of popular algorithms such as Stochastic Gradient Descent (SGD) that can be suited for different applications.
Most existing studies analyzing FL have been for the i.i.d. assumption of local user data. Some literature such as~\cite{wang2020optimizing} has studied the non-i.i.d. case with asynchronous communication reduction methods under privacy. However, extensive theoretical and application-oriented analysis of non-i.i.d. data-based FL remains to be explored. 
 

\section{Conclusion}
In this article, a forward-looking vision for 6G networks is outlined, highlighting specific use cases that extend or renew those introduced in 5G NR. 
Owing to the constraints inherent in wireless networks and performance specifications which will be especially stringent in 6G, distributed learning (DL) as a paradigm could play an important role in realizing novel applications. As a specific example, we apply Federated Learning (FL) with Multi-Agent Reinforcement Learning (MARL) to the Dynamic Spectrum Access (DSA) problem and demonstrate promising results through simulations. MARL-enabled FL is a good fit for 6G use cases that would rely on cooperation of large number of distributed users.
A relevant sampling of challenges and potential future directions for applying DL and FL approaches in 6G networks are also presented. 

\bibliographystyle{IEEEtran}
\bibliography{IEEEabrv,ref}

\end{document}